\documentclass[pre,aps,twocolumn,superscriptaddress,amsmath,showpacs,floatfix]{revtex4}
\usepackage{graphics}
\usepackage{amssymb}
\begin{document}
\title{The Effect of Thermal Fluctuations on Schulman Area Elasticity}

\author{O. Farago}
\email{farago@mrl.ucsb.edu}
\affiliation{
  Materials Research Laboratory, University of
  California, Santa Barbara, CA 93106}
\affiliation{
  Department of Physics, Korea Advanced Institute of
  Science and Technology (KAIST), 373-1 Kusong-dong,
  Yusong-gu, Taejon 305-701, South Korea.}
\author{P. Pincus}
\affiliation{
  Materials Research Laboratory, University of
  California, Santa Barbara, CA 93106}
\affiliation{
  Department of Physics, Korea Advanced Institute of
  Science and Technology (KAIST), 373-1 Kusong-dong,
  Yusong-gu, Taejon 305-701, South Korea.}
\date{\today}

\begin{abstract}
  \vspace{0.5cm} 
 
  We study the elastic properties of a two-dimensional fluctuating
  surface whose area density is allowed to deviate from its optimal
  (Schulman) value. The behavior of such a surface is determined by an
  interplay between the area-dependent elastic energy, the curvature
  elasticity, and the entropy. We identify three different elastic
  regimes depending on the ratio $A_p/A_s$ between the projected
  (frame) and the saturated areas. We show that thermal fluctuations
  modify the elastic energy of stretched surfaces ($A_p/A_s> 1$), and
  dominate the elastic energy of compressed surfaces ($A_p/A_s< 1$).
  When $A_p\sim A_s$ the elastic energy is not much affected by the
  fluctuations; the frame area at which the surface tension
  vanishes becomes smaller than $A_s$ and the area elasticity modulus
  increases.
\end{abstract}
\maketitle

\begin{widetext}
\section{Introduction}
\label{intro}

Surfactant molecules are formed by chemically bonding two parts with
different affinities to solvents - a hydrophilic head group (usually
ionic or polar), and a hydrophobic tail (usually composed of non-polar
hydrocarbon groups). When these molecules are placed in a mixture of
water and oil they tend to rest at the water-oil interface and
decrease the interfacial energy substantially \cite{rosen}. An
interface that is saturated by surfactants (as in the case of some
surfactant mixtures) has a {\em nearly-vanishing surface
  tension}\/\cite{degennes_taupin}. Other interfacial systems which
are frequently assumed to have zero surface tension are amphiphilic
fluid bilayers \cite{david_leibler}. Models ignoring surface tension
effects (and relating the physical properties to the curvature energy
only) have been very successful in describing many features of bilayer
systems, such as the shapes of vesicles \cite{seifert} and the
entropic interactions between membranes \cite{helfrichint}.

The problem with the notion of vanishing surface tension is that it
can be understood in two different ways, depending on the quantity in
question. To demonstrate the two possible meanings of the term
``surface tension'', we consider a {\em flat}\/ fluid membrane in an
aqueous solution. Let us first assume that the total number $N_s$ of
amphiphilic molecules forming the membrane is fixed, i.e., we ignore
the exchange of molecules between the surface (membrane) and the bulk
(solution) \cite{remarksurf}. This approximation may be relevant for
highly insoluble amphiphiles, such as lipids, whose {\em critical
  micelle concentration (cmc)}\/ is usually very low
\cite{lipowsky_sackmann}. (Typical values of the cmc for
bilayer-forming lipids are in the range $10^{-6}-10^{-10}$ M
\cite{israelachvili}). Lipid bilayers are also characterized by large
``residence times'' of the molecules within the aggregates
\cite{israelachvili,wimley}. The free energy of such a membrane
depends only on the surface density of the amphiphiles $\Sigma\equiv
N/A$, where $A$ is the area of the membrane.  It achieves a minimum at
a certain area density $\Sigma_0$ which is determined by the
intermolecular forces \cite{israelachvili,benshaul}.  Small deviations
from the equilibrium density can be described by a quadratic
approximation of the Helmholtz free energy
\begin{equation}
\Omega_C=\frac{K_A}{2A_s}\left(A-A_s\right)^2,
\label{elastic}
\end{equation}
where $A_s=N_s\Sigma_0$ is called the {\em saturated area}\/ of the
membrane, and $K_A$ is the stretching-compression modulus. Typical
values of $K_A$ for lipid bilayers are in the range $K_A\gtrsim 10^2\ 
{\rm ergs}/{\rm cm}^2$ \cite{israelachvili,evans_rawicz}. For such
values of $K_A$, any appreciable deviation of $A$ from $A_s$ would
involve an energy cost much larger than any other relevant energy
scale such as the thermal energy and the energy associated with
curvature deformations. Consequently, it is frequently assumed that
the total area is effectively constrained to $A_s$ and, therefore, the
{\em surface pressure}
\begin{equation}
\Pi \equiv -\frac{d\Omega_C}{dA}=0.
\end{equation}
It is the vanishing of $\Pi$ for $A=A_s$ which is usually described as
the vanishing of the surface tension.

The ideas presented above are essentially due to Schulman
\cite{schulman,degennes_taupin}, and so the saturated area is
sometimes called {\em Schulman area}. However, a careful examination
of Schulman's argument reveals that $\Pi$ (which we referred to as the
surface pressure) does not comply with our usual understanding of the
concept of surface tension. The quantity called surface tension is
supposed to describe the free energy required to remove molecules from
the bulk and create a unit area of the surface \cite{rowlinson_widom}.
It, therefore, cannot be discussed within a model that excludes the
exchange of molecules between the membrane and the solution (like the
one described above). This brings us to the second model: consider a
membrane whose area density is fixed to $\Sigma_0$, but whose total
number of molecules $N_s$ may vary. The grand potential $\Omega_G$ of
such a membrane can be described by
\begin{equation}
\Omega_G=\mu N_s=\mu\Sigma_0 A=\gamma A,
\label{surface}
\end{equation}
where the chemical potential $\mu$ is a Lagrange multiplier that fixes
the average number of surface molecules. From the above equation we
learn that (a) the quantity $\gamma$, which truly deserves the name
``surface tension'', is proportional to the chemical potential of the
molecules $\mu$; and (b) that there is no special reason for $\gamma$
to be equal to zero.

The differences between the above two examples can be viewed from a
slightly different perspective: In both cases we deal with saturated
surfaces. However, in the first example we deal with the response of
the membrane to the variation of the area density of the molecules.
The Helmholtz free energy $\Omega_C$ depicted by Eq.(\ref{elastic}) is
nothing but the {\em elastic free energy}\/ of the membrane, which
takes a quadratic form near the equilibrium (reference) state. In the
second case, on the other hand, we look at the process of creating
more saturated surface. The addition of molecules to the surface is
governed by the {\em interfacial free energy}\/ $\Omega_G$ described
by Eq.(\ref{surface}), which is linear in $A$.  Although we have just
argued that $\gamma$ and not $-\Pi$ should be regarded as the surface
tension of the system we will adopt, in what follows, a unified
approach: We will define the surface tension of a {\em flat} surface
as $\sigma\equiv dF/dA$, where $F$ (which we henceforth refer to as
the {\em free energy}\/) is the relevant thermodynamic potential,
i.e., $F=\Omega_C$ in the first case and $F=\Omega_G$ in the second.
(Note that $\Omega_C$ and $\Omega_G$ are {\em not}\/ related by a
Legendre transformation but rather describe two very distinct physical
problems.) The differentiation with respect to $A$ is taken while all
the other state variables are held constant \cite{remarkvar}.  We get
$\sigma=-\Pi$ when we apply our definition of the surface tension to
Eq.(\ref{elastic}), and $\sigma=\gamma$ when we apply it to
Eq.(\ref{surface}). For a {\em fluctuating}\/ surface we will use a
generalized version of the above definition for $\sigma$ [see
Eq.(\ref{deftension}), later in this section].

To fully describe the behavior of a thermally fluctuating surface we
need to include both the elastic free energy (\ref{elastic}) and the
interfacial free energy (\ref{surface}) in our Hamiltonian. The above
two models represent the two extreme cases of isolated elastic
surfaces (first model) and incompressible surfaces (second model).
There also exists a third model in which the surface is assumed to be
both isolated and incompressible \cite{seifert2,wortis,scott_safran}.
This last model has been successfully used in theoretical studies of
bilayer vesicles \cite{seifert}. As has been mentioned earlier, lipid
membranes are usually considered as nearly-incompressible due to the
separation of energy scales between the stretching and bending
energies. As a result, the behavior of bilayer systems had been hardly
studied in the framework of the first model, and very little is known
\cite{degennes_taupin,brochard,marsh} about the effect of thermal
fluctuations on the elastic free energy (\ref{elastic}) (which have
been described above for the case of a flat surface) and the
corresponding surface tension . This should be compared to our very
good understanding of the influence of temperature on the surface
tension $\gamma$ associated with incompressible surfaces
\cite{david_leibler,seifert2,helfrich_servuss,cai,morse-milner,
  feller_pastor,fournier}. The stretching elasticity contribution to
the surface tension is usually described by the linear relation
$\sigma=K_A(A-A_s)/A_s$, and is simply added as a correction to
$\gamma$ \cite{helfrich_servuss,evans_rawicz}.

The aim of the present paper is to fill the gap existing in the
literature and to study the effect of thermal fluctuations on Schulman
area elasticity. Understanding of this subject should improve our
analysis of the results of stretching experiments of water-insoluble
lipid bilayers \cite{evans_rawicz,seifert_lipowsky}. It is also
important for understanding the results of computer simulations where
frequently the number of molecules forming the surface is fixed
\cite{simulations}.  Most of the early work on this subject, which is
of a semiquantitative nature, is summarized in the review by de Gennes
and Taupin \cite{degennes_taupin}. More recently \cite{fournier},
there has been an attempt to introduce the stretching elasticity into
the problem in a non-ad hoc way, i.e., without simply adding it as a
correction to the surface tension $\gamma$. In Ref.~\cite{fournier}
the behavior of a membrane with a fixed microscopic area was
investigated, and the surface tension was associated with the
optically visible area. Here we consider membrane which is elastic at
the microscopic level and which is characterized by a finite
compressibility.

In this work we carry out a statistical mechanical analysis of the
behavior of a nearly flat fluid surface that spans a planar frame of a
total area $A_p$ (the ``projected area'') and consists of a fixed
number of molecules $N_s$. We will calculate the free energy of a
thermally fluctuating surface and extract the associated {\em frame
  tension}\/ which is defined by
\begin{equation}
\sigma\equiv \frac{dF}{dA_p}.
\label{deftension}
\end{equation}
The latter quantity gives the force per unit length experienced by the
frame holding the surface. The frame tension coincides with another
quantity (which is denoted by $r$) that has the dimensions of a
surface tension (i.e., energy per unit area) - the ``$q^2$
coefficient'' \cite{cai,farago,remarkcai}. The surface tension $r$ is the
one that is measured in flickering experiments
\cite{seifert_lipowsky}, where the dependence of the mean-square
amplitudes of the thermal fluctuations on the wavenumber $q$ is fitted
to the formula
\begin{equation}
\langle |u_q|^2\rangle = \frac{k_B T}
{A_p\left[rq^2+{\cal O}\left(q^4\right)\right]},
\label{meansquare}
\end{equation}
where $k_B$ is the Boltzmann constant and $T$ is the temperature. In
order to investigate the elastic behavior of the system we first
develop (in section \ref{thefunction}) a general formalism which can
be applied to any fluctuating surface whose energy can be written as
the sum of an area-dependent term and a curvature dependent term. The
validity of our formalism is demonstrated by applying it (in section
\ref{incompsurfaces}) to determine the surface tension of an
incompressible surfaces whose area-dependent energy is depicted by
Eq.(\ref{surface}). The well known result \cite{cai} for the
temperature dependent surface tension is recovered. We then turn (in
section \ref{compsurfaces}) to the main subject of the paper - the
elastic behavior of compressible fluctuating surfaces. Our discussion
of the subject is divided into three subsections: In subsection
\ref{flopsurfaces} we focus on the behavior of compressed surfaces
with the projected area $A_p$ smaller than the saturated (Schulman)
area $A_s$. We will try to see the extent to which thermal
fluctuations (which increase the total area compared to $A_p$) modify
the Schulman elastic energy. In subsection \ref{tensesurfaces} we
discuss the behavior of stretched surfaces with $A_p>A_s$. We will
examine whether the thermal fluctuations reduce or increase the (zero
temperature) elastic surface tension. The behavior of surfaces with
$A_p\sim A_s$ is studied in subsection \ref{schulsurfaces}. However,
we must stress here that for very small values of the surface tension
(which are encountered in part of this regime, close to the area where
$\sigma=0$) the elastic behavior is dominated by the interfacial free
energy (\ref{surface}) and the associated surface tension $\gamma$
rather than by the elastic free energy (\ref{elastic}). The case when
$A_p=A_s$ was investigated by Brochard et al.~\cite{brochard}. Their
interpretation of the surface tension had been reexamined
\cite{david_leibler} because it is based on the calculation of the
partition function at a single projected area. In section
\ref{summary} we summarize the results and discuss the other factors
that might influence the elastic behavior of the surface.

\section{The function $Z(\Delta A)$}
\label{thefunction}
  
We consider a nearly flat surface that spans a planar frame of a total
area $A_p$.  To describe the microscopic configurations of the surface
we use the Monge gauge $z=h(x,y)$, where $h$ is the height of the
surface above the frame reference plane \cite{remarkmonge}. We assume
that the Hamiltonian that describes the elastic energy of each
configuration can be decomposed into the sum of two terms:
\begin{equation}
{\cal H}={\cal H}_1\left(A\left[h\right]\right)
+\int_A dS\,\frac{1}{2}\kappa H^2.
\label{thehamiltonian}
\end{equation}
The first term on the right hand side (r.h.s) of the above equation
describes the dependence of the elastic energy on the total area $A$
of the surface and has yet to be specified. The second term is the
bending energy associated with the curvature. It describes the energy
difference between flat and curved surfaces with the same total area.
In Eq.(\ref{thehamiltonian}) the bending energy is expressed using the
quadratic approximation of the Canham-Helfrich Hamiltonian
\cite{canham,helfrich,remarkquad}, where $dS$ is a surface area
element, $H\equiv c_1+c_2$ is the sum of local principle curvatures
(the total curvature), and $\kappa$ is the associated bending modulus.
For simplicity we will restrict our discussion in this paper to
surfaces with no preferred (spontaneous) curvature, and to
fluctuations which do not change the topology of the surface. For a
nearly flat surface, i.e., when the derivatives of the height function
with respect to $x$ and $y$ are small -- $h_x, h_y\ll 1$, we have
\begin{equation}
A\simeq A_p+\int_{A_p} dxdy\,\frac{1}{2}\left(\nabla h\right)^2
\equiv A_p+\Delta A,
\label{areaapprox}
\end{equation}
and
\begin{equation}
\int_S dS\,H^2 \simeq 
\int_{A_p} dxdy\,\left(\nabla^2 h\right)^2,
\label{curvapprox}
\end{equation}
where the integral $\int_{A_p}$ runs over the frame reference surface,
and $\Delta A\geq 0$ [in Eq.(\ref{areaapprox})] denotes the excess
area of the surface due to the fluctuations in the normal $z$
direction.

Using Eqs.(\ref{thehamiltonian})--(\ref{curvapprox}), we write the
partition function of the system as
\begin{equation}
Z=\int{\cal D}\left[h\right]\exp
\left\{-\beta\left[{\cal H}_1\left(A_p+\int_{A_p} 
dxdy\,\frac{1}{2}(\nabla h)^2\right)+
\int_{A_p} dxdy\,\frac{1}{2}\kappa\left(\nabla^2 h\right)^2\right]
\right\},
\end{equation}
where $\beta=1/k_BT$. The above partition function can be also
written in the following form
\begin{eqnarray}
Z&=&\int_0^\infty d\left(\Delta A\right)\,
\exp\left[-\beta {\cal H}_1\left(A_p+\Delta A\right)\right]
\nonumber\\
&\times&
\int{\cal D}\left[h\right]\delta\left(\Delta A-\int_{A_p} 
dxdy\,\frac{1}{2}(\nabla h)^2\right)
\exp
\left[-\beta\kappa 
\int_{A_p} dxdy\,\frac{1}{2}\left(\nabla^2 h\right)^2\right]
\nonumber\\
&=&\int_0^\infty d\left(\Delta A\right)\,
\exp\left[-\beta {\cal H}_1\left(A_p+\Delta A\right)\right]
\times Z\left(\Delta A\right)
,
\label{za1}
\end{eqnarray}
where $\delta$ is the Dirac delta function. The function
$Z\left(\Delta A\right)$ can be identified as the partition function
of a surface whose total area is constrained to the value
$A=A_p+\Delta A$. In other words, it reflects the probability density
of the surface to have a total area $A$ when the different
configurations are weighted by their bending elasticity only. This
probability density is determined by two opposite trends: Entropy
favors strongly fluctuating configurations while the bending energy
makes the configurations with moderate slopes more preferable. The
contribution of the area-dependent term in the Hamiltonian
(\ref{thehamiltonian}) to the statistics is taken into account
separately by the additional Boltzmann factor, $\exp\left[-\beta {\cal
    H}_1\left(A_p+\Delta A\right)\right]$, in the integrand in
Eq.(\ref{za1}). In order to calculate the function $Z\left(\Delta
  A\right)$ we use the Fourier space representation of the delta
function
\begin{equation}
\delta(x-a)=\frac{1}{2\pi i}\int_{-i\infty}^{i\infty}
e^{\omega(x-a)}d\omega,
\end{equation}
which when substituted in Eq.(\ref{za1}) yields
\begin{eqnarray}
Z\left(\Delta A\right)&=&\frac{1}{2\pi i}
\int_{-i\infty}^{i\infty}d\omega\,e^{\omega\Delta A}
\nonumber\\
&\times&\int{\cal D}\left[h\right]
\exp
\left\{-\frac{1}{2}\int_{A_p} dxdy\,\left[
\omega\left(\nabla h\right)^2
+\beta\kappa \left(\nabla^2 h\right)^2\right]\right\}.
\label{za2}
\end{eqnarray}

So far we have treated our surface as if it were a smooth continuous
medium. One should not forget, however, that the Hamiltonian
(\ref{thehamiltonian}) is derived from a more microscopic description.
This means that the conformations of the surface cannot be defined
below some microscopic length scale: In the frame tangent plane we
must use a coarse-graining length $l$ comparable to the size of the
constituent molecules, so that an area element (``patch'') of linear
size $l$ contains at least one (preferably a few) molecules. Our
choice of $l$ (which is somewhat arbitrary) defines a coarse-graining
length scale $\lambda(l)$ for the normal displacements of the membrane
patches, where $\lambda(l)=\hbar\,(2\pi/mk_B T)^{1/2}$ is the thermal de
Broglie wavelength of a membrane patch, and $m$ in the definition of
$\lambda$ is the mass of the patch which is proportional to the (mean)
number of molecules forming it. The bending modulus $\kappa$ is also
scale-dependent \cite{safran,peliti_leibler}. 

Details on length scale smaller than $l$ are eliminated from our
statistical mechanical treatment of the system by considering only
those conformations of the surface $h(x,y)=h(\vec{r})$ with wave
vectors in the range $|q|<\Lambda\equiv2\sqrt{\pi}/l$:
\begin{equation}
h\left(\vec{r}\right)=\frac{\sqrt{A_p}}{\left(2\pi\right)^2}
\int_{|\vec{q}|\leq\Lambda}d\vec{q}\, l \lambda 
h_{q}e^{i\vec{q}\cdot\vec{r}}.   
\label{fourier}
\end{equation}
In the above equation $h_q$ is the amplitudes of the Fourier component
corresponding to $\vec{q}$ \cite{remarkunits}. The value of $\Lambda$
had been chosen to set the number of modes included in the spectrum to
be equal to the number of microscopic degrees of freedom $N$, i.e.,
\begin{equation}
N=\frac{A_p}{l^2}=\frac{A_p}{(2\pi)^2}\pi\Lambda^2.
\label{aqrelation}
\end{equation}
For simplicity we use a circular, rather than a square, region in
$q$-space. The sets of height functions $h(\vec{r})$ in real space
which can be generated by the modes included in the circular and
square Brillouin zones are different only in the details on the
microscopic length scale $l$. Therefore, both Brillouin zones are
equally adequate in describing the macroscopic behavior of the system.
With the Fourier modes representation, and using the identity
\begin{equation}
\int_{A_p}d\vec{r}\,e^{i(\vec{q}-\vec{p})\cdot\vec{r}}=
(2\pi)^2\,\delta(\vec{q}-\vec{p}),
\end{equation}
the excess area [Eq.(\ref{areaapprox})] is given by
\begin{equation}
\Delta A=\frac{A_p}{(2\pi)^2}\int_{\Lambda_0}^{\Lambda}
d\vec{q}\, \frac{1}{2}q^2l^2\lambda^2|h_q|^2,
\end{equation}
where $\Lambda_0=2\pi/\sqrt{A_p}\simeq 0$, and the mode $q=0$ has been
excluded as it corresponds to a constant shift in $h$ which does not
contribute to the excess area. The second integral in Eq.(\ref{za2})
now reads
\begin{equation}
\int{\cal D}\left[h_q\right]
\exp
\left\{-\frac{A_p}{(2\pi)^2}
\int_{\Lambda_0}^{\Lambda}d\vec{q}\, 
\frac{l^2\lambda^2}{2}\left[
\omega q^2+\beta\kappa q^4\right]|h_q|^2\right\}.
\label{fourierint}
\end{equation}
Tracing over the surface profile $h_q$ in Eq.(\ref{fourierint}), is
straightforward, giving
\begin{equation}
\exp\left\{
-\frac{A_p}{(2\pi)^2}\int_{\Lambda_0}^{\Lambda} \frac{d\vec{q}}{2}\,
\ln\left[\frac{\left(\omega q^2+\beta\kappa q^4\right)l^2\lambda^2}
{2\pi}\right]\right\}.
\end{equation}
This result should be substituted back in Eq.(\ref{za2}). The first
integral in Eq.(\ref{za2}) can be evaluated in the thermodynamic limit
using the method of steepest-descent. We find that (up to a constant
factor)
\begin{eqnarray}
Z\left(\Delta A\right)&\equiv&\exp\left\{\beta G\left(\Delta A\right)
\right\}\nonumber \\
\ &\simeq&
\exp\left\{\omega_s(\Delta A)\Delta A
-\frac{A_p}{(2\pi)^2}\int_{\Lambda_0}^{\Lambda}\frac{d\vec{q}}{2}\,
\ln\left[\frac{\left(\omega_s\left(\Delta A\right) q^2
+\beta\kappa q^4\right)l^2\lambda^2}
{2\pi}\right]\right\},
\label{za3}
\end{eqnarray}
where 
\begin{equation}
\omega_s=\beta\kappa\frac{\Lambda^2-\Lambda_0^2
\exp\left(8\pi\beta\kappa\Delta A/A_p\right)}
{\exp\left(8\pi\beta\kappa\Delta A/A_p\right)-1},
\label{omegaeq}
\end{equation}
satisfies
\begin{equation}
\frac{\partial G}{\partial \omega_s}=0,
\label{dgdomega}
\end{equation}
i.e., solves the equation
\begin{equation}
\Delta A-\frac{A_p}{(2\pi)^2}
\int_{\Lambda_0}^{\Lambda}\frac{d\vec{q}}
{2\left(\omega_s+\beta\kappa q^2\right)}=0.
\end{equation}

If we now substitute our expression for $Z\left(\Delta A\right)$
(\ref{za3}) in Eq.(\ref{za1}) and again use the saddle-point
approximation, we obtain
\begin{equation}
Z\simeq \exp\left\{\beta\left[G\left(\Delta A^*\right)-
{\cal H}_1\left(A_p+\Delta A^*\right)\right]\right\},
\end{equation}
where $\Delta A^*$ is the solution of the equation
\begin{equation}
-\frac{\partial {\cal H}_1}{\partial \Delta A}
+\frac{\partial G}{\partial \left(\Delta A\right)}
+\frac{\partial G}{\partial \omega_s}
\frac{\partial \omega_s}{\partial \left(\Delta A\right)}=0,
\end{equation}
and where the function $G\left(\Delta A,\omega_s\left(\Delta
    A\right)\right)$ is defined by Eq.(\ref{za3}). The last term in
the above equation vanishes by virtue of Eq.(\ref{dgdomega}), while
from Eq.(\ref{za3}) we have $\partial G/\partial \left(\Delta
  A\right)=\omega _s/\beta$.  We thus find that $\Delta A^*$ is the
solution of
\begin{equation}
\omega_s\left(\Delta A\right)=\beta \frac{\partial {\cal H}_1}
{\partial \left(\Delta A\right)}.
\label{deltaaequation}
\end{equation}
The free energy of the surface is given by
\begin{equation}
F=-k_BT\ln\left(Z\right)={\cal H}_1\left(A_p+\Delta A^*\right)
-G\left(\Delta A^*\right).
\label{fenergy}
\end{equation}
The first term in the above expression is the area-dependent energetic
contribution to the free energy. The second term can be regarded as
the entropic part of the free energy, where the different
configurations of the surface are weighted by their total bending
elasticity.

\section{Incompressible surfaces}
\label{incompsurfaces}

In order to evaluate the free energy of the surface we now need to
specify the area-dependent part ${\cal H}_1$ of the Hamiltonian
(\ref{thehamiltonian}). Let us first consider the extensively studied
example of an incompressible surface (see discussion in section
\ref{intro}). In this case, the area density of the surface is fixed,
and any change in the total area due to thermal fluctuations must be
matched by a change in the number of molecules that fixes the area per
molecule. The fluctuations in the area are governed by the Hamiltonian
[see Eq.(\ref{surface})]
\begin{equation}
{\cal H}_1=\gamma A=\gamma\left(A_p+\Delta A\right),
\label{h1incompressible}
\end{equation}
where (as discussed in section \ref{intro}) $\gamma$ is directly
proportional to the chemical potential for the addition of molecules
to the surface. From Eqs.(\ref{deltaaequation}) and
(\ref{h1incompressible}) we find that
\begin{equation}
\omega_s=\beta \gamma.
\label{deltaaequation1}
\end{equation}
Using Eqs.(\ref{za3}), (\ref{fenergy}), and (\ref{deltaaequation1}) we
obtain the expression for the free energy (consistent with \cite{cai})
\begin{eqnarray}
F&=&
\gamma A_p+k_BT\frac{A_p}{(2\pi)^2}\int_{\Lambda_0}^{\Lambda} 
\frac{d\vec{q}}{2}\,\ln\left[\frac{\left(\beta\gamma q^2
+\beta\kappa q^4\right)l^2\lambda^2}
{2\pi}\right] 
\nonumber \\
\ &=& A_p\left\{\gamma+\frac{k_BT}{(2\pi)^2}
\int_{\Lambda_0}^{\Lambda}\frac{d\vec{q}}{2}\,
\ln\left[\frac{\left(\beta\gamma q^2
+\beta\kappa q^4\right)l^2\lambda^2}{2\pi}\right] \right\},
\label{fenergy1}
\end{eqnarray}
which is correct to the lowest order in an expansion in
$(\beta\kappa)^{-1}=k_BT/\kappa$. To find the surface tension, $\sigma$, we
need to take the full derivative of $F$ with respect to $A_p$.  It is
therefore important to examine the implicit dependence on $A_p$ of the
expression inside the braces in Eq.(\ref{fenergy1}). One part of this
expression that obviously depends on $A_p$ is the lower limit of the
integral $\Lambda_0=2\pi/\sqrt{A_p}$. This dependence is associated
with the logarithmic finite size correction to the free energy, whose
origin is the fact the variation of the linear size of the frame
($\sqrt{A_p}$) leads to changes in the wave-numbers of the long
wavelength modes. The contribution of this effect to $\sigma$ is
negligible in the thermodynamic limit. A more subtle issue is the
possible dependence on $A_p$ of the microscopic length $l$ and of the
quantities $\Lambda$ and $\lambda$ which are directly related to $l$.
Recall that the patches have been identified as small sections of the
surface containing, on average, a given number of (at least one)
molecules.  Since the area density of the surface is fixed, $l$ can be
derived from the constraint
\begin{equation}
\frac{\langle A\rangle}{N}=l^2\frac{\langle A \rangle}{A_p}={\rm Const.},
\label{criterion1}
\end{equation}
where $\langle A \rangle$ is the mean total area of the surface. Using
Eq.(\ref{fenergy1}) and the relation $\langle A \rangle=\partial
F/\partial \gamma$, we get \cite{helfrich_servuss}
\begin{equation}
\langle A \rangle =A_p\left\{1+\frac{1}{8\pi\beta\kappa}
\ln\left[\frac{\gamma+\kappa\Lambda^2}{\gamma+\kappa\Lambda_0^2}
\right]\right\}.
\label{incompelaw}
\end{equation}
For $\gamma\neq 0$ and in the limit $A_p\rightarrow \infty$ (
$\Lambda_0=2\pi/\sqrt{A_p}\rightarrow0$), we readily conclude that
criterion (\ref{criterion1}) is obeyed by setting $l$ to a constant
value which does not depend on $A_p$. We thus find (ignoring the above
mentioned finite size correction) in agreement with \cite{cai} that
\begin{equation}
\sigma=\frac{dF}{dA_p}\simeq \frac{F}{A_p}=
\gamma+\frac{k_BT}{(2\pi)^2}
\int_{\Lambda_0}^{\Lambda}\frac{d\vec{q}}{2}\,
\ln\left[\frac{\left(\beta\gamma q^2
+\beta\kappa q^4\right)l^2\lambda^2}{2\pi}\right].
\label{gammaincomp}
\end{equation}
If $\gamma=0$ we find $l$ depending on $A_p$, what brings in another
finite size correction to the above result (\ref{gammaincomp}).

\section{Compressible surfaces}
\label{compsurfaces}
We now turn to study a surface with a {\em fixed}\/ number of
molecules, whose area elasticity can be approximated by the harmonic
form [see Eq.(\ref{elastic})]
\begin{equation}
{\cal H}_1=\frac{K_A}{2A_s}\left(A-A_s\right)^2=\frac{K_A}{2A_s}
\left(\Delta A-\Delta A_s\right)^2,
\label{schulmanenergy}
\end{equation}
where $\Delta A_s\equiv A_s-A_p$.  The system under consideration is
characterized by three energy scales: $k_BT=(\beta)^{-1}$, $\kappa$,
and $K_Aa_s$, where $a_s\equiv A_s/N$ is the Schulman area of the
surface patches (the microscopic degrees of freedom). In what follows
we will assume that 
\begin{equation}
\sqrt{\beta K_A a_s}\gg \beta \kappa\gg 1.
\label{nearlyincomp}
\end{equation}
These relations are obeyed by typical values of the elastic moduli and
the Schulman (saturated) area per lipid of phospholipid bilayers. We are
interested in calculating the free energy of such a surface as a
function of the ratio $A_p/A_s$ between the projected and the
saturated areas. Combining Eqs.(\ref{omegaeq}),
(\ref{deltaaequation}), and (\ref{schulmanenergy}) we arrive to the
following equation
\begin{equation}
\omega_s\left(\Delta A\right)=
\beta\kappa\frac{\Lambda^2-\Lambda_0^2
\exp\left(8\pi\beta\kappa\Delta A/A_p\right)}
{\exp\left(8\pi\beta\kappa\Delta A/A_p\right)-1}=
\beta\frac{K_A}{A_s}\left(\Delta A-\Delta A_s\right),
\label{theequation}
\end{equation}
whose solution $\Delta A^*$ should be substituted in expression
(\ref{fenergy}) for the free energy of the surface. Three regimes can
be distinguished:

\subsection{Floppy surfaces} 
\label{flopsurfaces}

Consider the case when $A_p<A_s$. The elastic energy caused by the
mismatch between the frame area and the saturated area can be relieved
by thermal fluctuations which store the extra area needed to bring the
total area $A$ close to $A_s$. Let $A_p^*$ be the projected area for
which the solution of Eq.(\ref{theequation}) coincides with Schulman
area
\begin{equation}
A^*=A_s,
\label{areazero}
\end{equation}
so that 
\begin{equation}
\omega_s=0.
\label{omegazero}
\end{equation}
One can easily verify that
\begin{equation}
A_p^*= A_s
\left[1+\frac{1}{8\pi\beta\kappa}\ln\left(\frac{\Lambda^2}{\Lambda_0^2}\right)
\right]^{-1}=
A_s\left[1+\frac{1}{8\pi\beta\kappa}\ln\left(\frac{N}{\pi}\right)
\right]^{-1}.
\end{equation}
For $A_p\sim A_p^*$ the corrections to Eqs.(\ref{areazero}) and
(\ref{omegazero}) can be expressed as a power series in the variable
$\epsilon_A\equiv A_p-A_p^*$ ($|\epsilon_A|<[{\rm max}(\ln
N,8\pi\beta\kappa)]^{-1}A_p^*$). The following relations can be
derived:
\begin{equation}
A^*=A_s+\epsilon_A\frac{(2\pi)^2\kappa A_s}
{K_A {A_p^*}^{\, 2}}
\left[\ln\left(\frac{N}{\pi}\right)+1-8\pi\beta\kappa
+{\cal O}\left(\frac{\epsilon_A}{A_P^*}\right)\right]
,
\label{areaone}
\end{equation}
and 
\begin{equation}
\omega_s=\frac{(2\pi)^2\beta\kappa}
{{A_p^*}}\left[\ln\left(\frac{N}{\pi}\right)+1-8\pi\beta\kappa\right]
\frac{\epsilon_A}{A_p^*}+\cdots\ ,
\end{equation}
which when used together with Eqs.(\ref{za3}), (\ref{fenergy}), and
(\ref{schulmanenergy}), yield (after some straightforward, but lengthy
calculation) the following expression for the free energy
\begin{eqnarray}
F&=&F_0+F_1+\cdots=k_BT\frac{A_p}{(2\pi)^2}\int_{\Lambda_0}^{\Lambda} 
\frac{d\vec{q}}{2}\,\ln\left[\frac{\beta\kappa q^4l^2\lambda^2}
{2\pi}\right]
\nonumber \\
\ &+&
k_BT
\left[\ln\left(\frac{N}{\pi}\right)+1-8\pi\beta\kappa\right]
\left [\frac{3}{8\pi}\ln\left(\frac{N}{\pi}\right)+\frac{\pi}{2}\right]
\frac{\epsilon_A}{A_p^*}+\cdots.
\label{fenergyflop}
\end{eqnarray}
The form of the first term $F_0$ in free energy (\ref{fenergyflop}) is
similar to (\ref{fenergy1}) for an incompressible surface with
$\gamma=0$. The second term $F_1$ is the leading linear correction in
the small variable $\epsilon_A/A_p^*$. The harmonic (Schulman) elastic energy
contributes only to the quadratic correction.

Our result (\ref{fenergyflop}) should be compared with the free energy
(\ref{gammaincomp}) of an incompressible surface in the limit
$\gamma\rightarrow 0$. As has been discussed in the literature
\cite{david_leibler} (see also our discussion in section \ref{intro})
and as evident from our Eq.(\ref{gammaincomp}), the vanishing of
$\gamma$ should not be confused with the vanishing of surface tension
$\sigma$ which occurs when $A_p\sim A_s$ (as we show in section
\ref{schulsurfaces}). In the derivation of Eq.(\ref{gammaincomp}),
which applies for the case of an incompressible surface and for which
the total number of surface molecules is {\em not}\/ fixed, we have
used relation (\ref{criterion1}), which enforces the conservation of
mass of the surface patches. Here, we consider a different scenario,
where the total number of molecules is fixed. Consequently, relation
(\ref{criterion1}) should be replaced with Eq.(\ref{aqrelation}) in
which we fix $N$, the number of patches dividing the surface. In order
to calculate $\sigma$ by differentiating $F$ with respect to $A_p$, it
is better to consider the explicit expression for $F_0$ and to perform
the integral in Eq.(\ref{fenergyflop}). Setting the lower limit
$\Lambda_0=0$ we get
\begin{equation}
F_0\simeq k_BT\frac{A_p}{4\pi}\Lambda^2\left[\ln\left(
\frac{\beta\kappa\lambda^2l^2\Lambda^4}{2\pi}\right)-2\right]=
-Nk_BT\left[\ln\left(
\frac{A_p}{8\pi\beta\kappa\lambda^2N}\right)+2\right].
\label{fenergyflop2}
\end{equation}
Apart from the constant $-2Nk_BT$ which can be discarded (if
temperature is fixed), the free energy depicted by the above equation
is reminiscent of the free energy
$F=-Nk_BT\ln\left(A_p/\lambda^2N\right)$ of a two-dimensional ideal
gas consisting of $N$ particles confined in an area $A_p$. The only
difference between the free energies of those two systems is that the
{\em effective}\/ de Broglie thermal wavelength in
Eq.(\ref{fenergyflop2}) is $\lambda^*=\lambda\sqrt{8\pi\beta\kappa}$.
While $\lambda$ is a constant that depends only on the mean number of
molecules included in each patch (which is not affected by changes in
$A_p$ if we use Eq.(\ref{aqrelation}) with a fixed value of $N$), the
effective thermal wavelength $\lambda^*$ may be area-dependent. The
origin of this dependence is the bending modulus $\kappa$ appearing in
the expression for $\lambda^*$. In the quadratic approximation of the
Canham-Helfrich Hamiltonian [Eq.(\ref{thehamiltonian})], the elastic
bending energy is expanded around the flat reference state of area
$A_p$ and, in general, $\kappa$ depends on the properties of the
reference state.  As the number of surface molecules is fixed, the
variation of $A_p$ leads to the variation of the area density of the
reference state and, presumably, to the variation of $\kappa$. The
exact calculation of $\kappa$ requires the knowledge of the molecular
interactions between the molecules \cite{blokhuis,farago}, and is
beyond the scope of this paper.  We will, therefore, restrict our
discussion in the reminder of this paper to surfaces for which the
variation of $\kappa$ (over the range of $A_p$ of interest) is
negligible. In such a case,
\begin{equation}
\sigma_0\equiv\frac{dF_0}{dA_p}\simeq -\frac{Nk_BT}{A_p}.
\end{equation}
The first correction to this result is
\begin{equation}
\sigma_1\equiv\frac{dF_1}{dA_p}=\frac{dF_1}{d\epsilon_A}=
\left[\ln\left(\frac{N}{\pi}\right)+1-8\pi\beta\kappa\right]
\left [\frac{3}{8\pi}\ln\left(\frac{N}{\pi}\right)+\frac{\pi}{2}\right]
\frac{k_BT}{A_p^*}.
\end{equation}
In the thermodynamic limit this is merely a logarithmic size correction to
$\sigma_0$, and we thus conclude that for $A_p\sim A_p^*$
\begin{equation}
\sigma\simeq -\frac{Nk_BT}{A_p}.
\label{negstension}
\end{equation}
The negative surface tension implies that the surface opposes its
contraction. This result is expected since reduction of $A_p$ requires
the surface to bend more and as a consequence pay a higher energy cost
in order to attain the Schulman area. Negative surface tension with
the same origin was previously explained in the context of vesicles in
Ref.~\cite{seifert2}, and has been also observed in simulations of
bilayer membranes \cite{farago2,goetz,marrink_mark}.  While the very
fact that the surface tension becomes negative is not surprising, its
magnitude ($k_BT$ per microscopic unit area, i.e., of the order of
1--10 ${\rm ergs}/{\rm cm}^2$), as predicted by Eq.(\ref{negstension}),
is strikingly large. Such a large and negative surface tension implies
that undulations with wavelengths longer than
$2\pi\sqrt{\beta\kappa(A_p/N)}$ should be unstable, because the
coefficients of $|h_q|^2$ in the free energy become negative. This
undulation instability will not show up in computer simulations where
the size of the sample is small \cite{farago2,goetz,marrink_mark}. For
real physical systems this highly compressed regime is unattainable.
It can be preempted by a fluid-solid phase transition \cite{farago2}
(in which case the shear modulus of the solid surface may act against
the increase in the total area), or by a transition of molecules from
the surface to the solution \cite{milner}. It should be noted here
that a large negative surface tension of the same order of magnitude
is also predicted by the incompressible surface model [see
Eq.(\ref{gammaincomp}), with $\gamma=0$]. However, it is probably
worthwhile reemphasizing that surfaces are nearly-tensionless for
$\sigma\sim 0$ rather than for $\gamma\sim0$.

\subsection{Tense surfaces}
\label{tensesurfaces}

Consider the case when $A_p>A_s$. In this regime the saturated area is
not attainable because $A>A_p$. For
\begin{equation}
\frac{A_p}{A_s}-1\gg \sqrt{\frac{1}{2\beta K_Aa_s}}
\end{equation}
(a condition which for $\sqrt{\beta K_Aa_s}\gg1$ excludes only a small regime
of $A_p$ close to $A_s$), i.e. for $\Delta A_s\ll -A_s\left(2\beta
  K_Aa_s\right)^{-1/2}$, the optimal area is very close to $A_p$ so
that Eq.(\ref{theequation}) can be approximated by
\begin{equation}
\omega_s\left(\Delta A\right)=
\frac{\Lambda^2A_p}{8\pi\Delta A}=-\beta\frac{K_A}{A_s}\Delta A_s.
\label{daequationtense}
\end{equation} 
The above approximation is obtained by neglecting the exponent in the
numerator of the expression in the central part of
Eq.(\ref{theequation}), and expanding the exponent in the denominator
close to $\Delta A=0$.  We also omit the term $\Delta A$ on the r.h.s
of Eq.(\ref{theequation}). The solution of Eq.(\ref{daequationtense})
is
\begin{equation}
\Delta A^*=-\frac{A_sA_p\Lambda^2}{8\pi\beta K_A\Delta A_s}=
-\frac{NA_s}{2\beta K_A\Delta A_s}=
-\frac{A_s^2}{2\beta K_Aa_s\Delta A_s}.
\label{solutiontesne}
\end{equation}
(Note that for the range of $\Delta A_s$ considered here the solution
satisfies $\Delta A^*\ll|\Delta A_s|$, and so our approximation of the
r.h.s of Eq.(\ref{theequation}) is justified.). Combining
Eqs.(\ref{za3}), (\ref{fenergy}), (\ref{schulmanenergy}),
(\ref{daequationtense}), and (\ref{solutiontesne}), and recalling that
$\Delta A^*\ll|\Delta A_s|$, we find that the free energy is given by
\begin{eqnarray}
F&\simeq& \frac{K_A}{2A_s}(\Delta A_s)^2-
\frac{Nk_BT}{2}
\nonumber \\
\ &+&k_BT\frac{A_p}{(2\pi)^2}\int_{\Lambda_0}^{\Lambda} 
\frac{d\vec{q}}{2}\,\ln\left[\frac{\left(\beta 
\left(K_A|\Delta A_s|/A_s\right)q^2
+\beta\kappa q^4\right)l^2\lambda^2}
{2\pi}\right].
\label{fenergytense}
\end{eqnarray}
If temperature is fixed then the second term in the above equation is
a constant and can be ignored. The first term depicts the energetic
contribution of Schulman elasticity to $F$. This is simply the elastic
energy of a flat surface having been stretched to an area $A_p>A_s$.
The surface tension experienced by such a flat surface is
\begin{equation}
\sigma_0=-\frac{K_A}{A_s}\Delta A_s=\frac{K_A}{A_s}|\Delta A_s|>0.
\label{linear}
\end{equation}
The third term in Eq.(\ref{fenergytense}) represents the entropic part
of $F$. It is similar to the second term in Eq.(\ref{fenergy1}), which
is the leading thermal correction to the free energy of an
incompressible surface, with $\gamma=\sigma_0$. The (full) derivative
of this term with respect to $A_p$ yields the correction to the linear
relation $\sigma=\sigma_0$, between the surface tension and $\Delta
A_s$. It can be shown that for $K_Aa_s/\kappa\gg 1$ [see relation
(\ref{nearlyincomp})], the sign of this correction is positive which
means that thermal fluctuations lead to a super-Hookean elasticity.
This observation can be understood by noting [see
Eq.(\ref{solutiontesne})] that the optimal area $A^*$ is independent
of $\kappa$, and that it gets closer to $A_p$ by increasing $|\Delta
A_s|$. The approach of $A^*$ to $A_p$ involves a strong suppression of
the thermal fluctuations, and that naturally reduces the entropy of
the surface and thereby increases the surface tension. As already
said, this entropic surface tension is also found for incompressible
surfaces \cite{seifert2,helfrich_servuss,cai,morse-milner,
  feller_pastor,fournier}. Stretching experiments of vesicles
\cite{evans_rawicz} are often analyzed assuming the decoupling of the
elastic and entropic surface tensions. However, our
Eq.(\ref{fenergytense}) suggests the existence of a rather complicated
interplay between the elastic and the entropic contributions, which is
expressed by the fact that the stretching modulus $K_A$ appears in the
integral expression for the entropic component.

\subsection{Schulman surfaces}
\label{schulsurfaces}

Finally, we discuss the case when $A_p\sim A_s$. More
precisely, we consider the regime where the projected and saturated
areas are sufficiently close to each other that
\begin{equation}
|\Delta A_s|< \sqrt{\frac{1}{2\beta K_Aa_s}}\,A_s\equiv\sqrt{D} A_s
\ll A_s,
\label{schullimit}
\end{equation}
where $\Delta A_s$ is either positive or negative. Instead of
Eq.(\ref{daequationtense}) we now have
\begin{equation}
\omega_s\left(\Delta A\right)=
\frac{\Lambda^2A_p}{8\pi\Delta A}=\beta\frac{K_A}{A_s}
\left(\Delta A-\Delta A_s\right).
\label{daequationschul}
\end{equation}
Up to second order in the parameter $\Delta A_s/(\sqrt{D}A_s)<1$,
the solution of this equation is
\begin{equation}
\Delta A^*=\sqrt{D}A_s+\frac{\Delta A_s}{2}+\frac{\left(\Delta A_s\right)^2}
{8\sqrt{D}A_s},
\label{solutionschul1}
\end{equation}
and hence
\begin{equation}
\omega_s\left(\Delta A^*\right)=
\beta K_A\left(\sqrt{D}-\frac{\Delta A_s}{2A_s}+
\frac{\left(\Delta A_s\right)^2}
{8\sqrt{D}A_s^2}\right).
\label{solutionschul2}
\end{equation}
Eqs.(\ref{za3}), (\ref{fenergy}), (\ref{solutionschul1}), and
(\ref{solutionschul2}) yield the following form for the free energy
\begin{eqnarray}
&F&\simeq \frac{K_A}{2A_s}(\Delta A_s)^2-DK_AA_s
\\
&+&k_BT\frac{A_p}{(2\pi)^2}\int_{\Lambda_0}^{\Lambda} 
\frac{d\vec{q}}{2}\,\ln\left[\frac{\beta l^2\lambda^2q^2}
{2\pi}\left(K_A\sqrt{D}-\frac{K_A\Delta A_s}{2A_s}+
\frac{K_A\left(\Delta A_s\right)^2}
{8\sqrt{D}A_s^2}+\kappa q^2\right)\right].
\nonumber
\label{fenergyschul}
\end{eqnarray}
The last three terms in the argument of the logarithmic function are
smaller than the first one [see relations (\ref{nearlyincomp}) and
(\ref{schullimit})]. Using a Taylor expansion of the logarithm and
keeping only terms up to first order in $\beta\kappa/\sqrt{\beta K_A
  A_s}$ and second order in $\Delta A_s/(\sqrt{D}A_s)$ we obtain
\begin{eqnarray}
&F&\simeq \frac{K_A}{2A_s}(\Delta A_s)^2-DK_AA_s
+k_BT\frac{A_p}{(2\pi)^2}\int_{0}^{\Lambda} 
\frac{d\vec{q}}{2}\,\ln\left(\frac{\beta l^2\lambda^2q^2}
{2\pi}K_A\sqrt{D}\right)
\nonumber \\
&+&k_BT\frac{A_p}{(2\pi)^2}\int_{0}^{\Lambda} 
\frac{d\vec{q}}{2}\,
\left[ \frac{{\displaystyle \frac{-K_A\Delta A_s}{2A_s}}
+{\displaystyle \frac{K_A\left(\Delta A_s\right)^2}{8\sqrt{D}A_s^2}}}
{K_A\sqrt{D}}
+\frac{-{\displaystyle \frac{\left(K_A\Delta A_s\right)^2}{8A_s^2}}}
{K_A^2D}
+\frac{\kappa q^2}{K_A\sqrt{D}}
\right]
\nonumber \\ 
\ &=&
\frac{K_A}{2A_s}(\Delta A_s)^2-DK_AA_s+Nk_BT
\left[-1+\ln\left(2\beta K_A\lambda^2\sqrt{D}\right)\right]
\nonumber \\
&+&k_BT\frac{A_p}{(2\pi)^2}\int_{0}^{\Lambda} 
\frac{d\vec{q}}{2}\,
\left[ \frac{{\displaystyle \frac{-K_A\Delta A_s}{2A_s}}
+{\displaystyle \frac{K_A\left(\Delta A_s\right)^2}{8\sqrt{D}A_s^2}}}
{K_A\sqrt{D}}
+\frac{-{\displaystyle \frac{\left(K_A\Delta A_s\right)^2}{8A_s^2}}}
{K_A^2D}
+\frac{\kappa q^2}{K_A\sqrt{D}}
\right].
\label{fenergyschu2}
\end{eqnarray}
If we fix the temperature and ignore the small dependence of $\kappa$
on $A_p$, then the second and third terms in Eq.(\ref{fenergyschu2})
are constants which can be dismissed. The first term is the
zero-temperature Schulman elastic energy, while the forth term is the
thermal correction. The integrals in the latter can be easily
performed, and after several mathematical manipulations we find that
(up to an irrelevant additive constant)
\begin{equation}
F\simeq \frac{K_A}{2A_s}(\Delta A_s)^2\left(1+\frac{16\pi\kappa\sqrt{D}}
{K_Aa_s}\right)
+K_A\sqrt{D}\Delta A_s\left(-\frac{1}{2}+\frac{8\pi\kappa}{K_Aa_s}\right).
\label{fenergyschu3}
\end{equation}
In the $\Delta A_s$ regime which we discuss in this section [see
Eq.(\ref{schullimit})] the second term in Eq.(\ref{fenergyschu3})
should not be considered as correction to the first, but rather as the
dominant term in the free energy. From Eq.(\ref{fenergyschu3}) we
conclude that thermal fluctuations leads to a shift in the value of
the projected area at which the free energy is minimal. The free
energy attains its minimum at
\begin{equation}
A_p=A_s\left\{1-\sqrt{D}\left[\frac{1}{2}
+{\cal O}\left(\frac{\kappa}{K_Aa_s}\right)\right]\right\}\equiv A_s^{\rm eff},
\end{equation}
which can be regarded as the effective Schulman area at which
$\sigma=0$. The very fact that $A_s^{\rm eff}<A_s$ is not surprising
since the total area is always larger than the projected area. For
typical values of phospholipids: $\kappa=10k_BT\sim 5\times 10^{-13}$
ergs, $K_A\sim 200$ ergs/${\rm cm}^2$, and $a_s\sim 10^{-14}$ ${\rm
  cm}^2$ (the area of a patch consisting of 1-2 molecules), we find
that $\sqrt{D}\sim 0.1$ and so the shift in the equilibrium projected
area is about $5\%$ of $A_s$. We also conclude from
Eq.(\ref{fenergyschu3}) that thermal fluctuations increase the
compression-stretching modulus $K_A$. The effective modulus is
\begin{equation}
K_A^{\rm eff}\simeq K_A+\frac{16\pi\kappa\sqrt{D}}{a_s}.
\end{equation}
For the values of the relevant quantities quoted above
(characteristic of phospholipid bilayers) we find, quite surprisingly,
that the magnitude of the thermal stretching modulus is comparable to
the bare modulus $K_A$.

In the Evans-Rawicz experiment \cite{evans_rawicz}, linear elastic
response was measured for areal strain $\eta\equiv (A_p-A_s^{\rm
  eff})/A_s^{\rm eff}<0.05$. This is precisely the regime studied in
the present section, and so our theoretical discussion explains well
the experimental behavior. For extremely small values of the $\eta$
($<0.005$) the surface tension grows exponentially with the strain.
This elastic behavior has been recently explained in the framework of
the incompressible surface model, and has been attributed to the
dependence of the entropic surface tension on the optically visible
area \cite{fournier}. 

\section{Summary}
\label{summary}

We have studied the elasticity of fluctuating two-dimensional
compressible (elastic) surfaces. Starting with a Hamiltonian including
the Canham-Helfrich bending energy and an area-dependent term, we have
first developed a general formalism for analyzing the
statistical-mechanical behavior of such systems. We have used the
formalism to reproduce the expression for the temperature-dependent
tension of incompressible surfaces. Our investigation of the elastic
behavior of compressible surfaces reveals that the minimum of the free
energy is obtained when the projected (frame) area $A_p=A_s^{\rm
  eff}$, where $A_s^{\rm eff}$ is slightly smaller than the saturated
area $A_s$. In the vicinity of $A_s^{\rm eff}$ the elastic energy is
depicted by the Schulman quadratic form (\ref{schulmanenergy}) with an
area elasticity modulus $K_A^{\rm eff}$ slightly larger than the bare
modulus $K_A$. The response of the surface to compression and
stretching away from $A_s^{\rm eff}$ is very different. The stretching
behavior is dominated by the area elasticity. Thermal fluctuations
introduce a correction to Hooke's law, which increases the surface
tension. By contrast, area elasticity plays a very small role in the
compression behavior because the total area of the fluctuating surface
is (effectively) constrained to $A_s$.  The negative surface tension
observed in this regime should be attributed to fact that upon
compression, the total area stored by the fluctuations (the excess
area) grows. The enhancement of thermal fluctuations leads to an
increase in the bending energy of the surface.

We conclude by briefly reviewing the factors which have been left out
of our model, and which influence the elastic behavior of the surface.
One of them is the exchange of molecules between the surface and the
solution. It can be dealt with within the framework of a
grand-canonical ensemble where both the area $A$ and the number of
patches $N$ are allowed to fluctuate. Understanding the dependence of
the bending modulus $\kappa$ on the density of the reference state
(i.e., on the ratio $A_p/A_s$) is another challenge. This is a
complicated matter since the values of the phenomenological parameters
$\kappa$, $A_s$, and $K_A$ are determined by the structure of the
molecules and the interactions between them. It would therefore
require a theoretical study at the molecular level (see, e.g.
\cite{benshaul,farago2,blokhuis}). For strongly fluctuating surfaces we should
include corrections to the quadratic form of the Canham-Helfrich
Hamiltonian and to go beyond the lowest order in $k_BT/\kappa$. In
that case ``measure factors'' \cite{cai} which rather complicate the
mathematical treatment must be introduced in order to correct the
partition function.  Finally, we mention the possibility that
strongly-stretched membranes can reduce their elastic energy by
developing pores. A discussion at this subject can be found elsewhere
\cite{farago2,litster,netz,muller,sens_safran,schillcock}.

{\em Acknowledgments:}\/ We thank Thomas Powers, Jean-Baptiste
Fournier, and Armand Ajdari for helpful discussions. This work was
supported by the National Science Foundation under Award
No.~DMR-0203755. The Materials Research Laboratory at UC Santa Barbara
is supported by NSF No.~DMR-0080034. 


\end{widetext}


\begin{thebibliography}{99}
\bibitem{remarksurf} The term ``surface'' is used in this paper to
  describe a variety of quasi two-dimensional systems including simple
  interfaces, Langmuir monolayers, and self-assembled bilayer
  membranes.
\bibitem{remarkcai} In Ref.~{\protect \cite{cai}} the authors derive
  the relation $r=F/A_p$ which is correct only for incompressible
  surfaces. In Ref.~{\protect \cite{farago}} this result is discussed
  in the context of compressible surfaces, and it is demonstrated that
  the correct form (which applies for both compressible and
  incompressible surfaces) should be $r=dF/dA_p$.
\bibitem{remarkmonge} Since the Monge representation describes the
  conformation of the surface using a single-valued height function,
  it excludes configurations in which the surface forms overhangs. In
  addition to this approximation we also assume that the height
  function has a moderate slope [see {\protect Eqs.(\ref{areaapprox})
    and (\ref{curvapprox})}].
\bibitem{remarkquad} {\protect The expression for the curvature energy
    in Eq.(\ref{thehamiltonian}) is the leading term in an expansion
    of the elastic energy for small curvatures.}
\bibitem{remarkunits} The quantities {\protect $h_q$} and {\protect
    $u_q$} appearing in Eqs.({\protect \ref{fourier}}) and ({\protect
    \ref{meansquare}}) respectively are proportional to each other.
  They both describe the amplitudes of the modes in the Fourier
  transform of {\protect $h(\vec{r})$}, but using different units.
  The former is dimensionless while the latter has the dimensions of
  length.
\end{thebibliography}
\end{document}